\documentclass[12pt]{iopart}
\usepackage{iopams}  
\usepackage[utf8]{inputenc}
\usepackage{graphicx}
\usepackage{dcolumn}
\usepackage{bm}
\usepackage{enumitem}
\usepackage{siunitx}
\usepackage{booktabs}
\usepackage{hyperref}
\hypersetup{
    colorlinks=true,
    linkcolor=blue,
    filecolor=magenta,      
    urlcolor=cyan
    }
\usepackage{amsfonts}
\usepackage{nicefrac}
\usepackage{microtype}
\usepackage{lipsum}
\usepackage{multirow}
\usepackage{color}
\usepackage[normalem]{ulem}
\usepackage{placeins}
\usepackage{xspace}
\usepackage{soul}
\usepackage{listing}
\usepackage[para]{threeparttable}

\usepackage[pagewise]{lineno}

\newcommand{\units}[1]{\ensuremath{\text{\,#1}}\xspace}

\newcommand{\ptmomentum}{\ensuremath{p_{\mathrm{T}}}\xspace}
\newcommand{\GeV}{\ensuremath{\,\text{Ge\hspace{-.08em}V}}\xspace}
\newcommand{\TeV}{\ensuremath{\,\text{Te\hspace{-.08em}V}}\xspace}
\newcommand{\hlsfml}{\texttt{hls4ml}\xspace}

\begin{document}

\begin{flushright}
    \footnotesize
    FERMILAB-PUB-24-0030-CMS-CSAID-PPD
\end{flushright}

\title[Ultrafast Jet Classification]{Ultrafast jet classification at the HL-LHC}

\author{Patrick Odagiu$^{1}$, Zhiqiang Que$^2$,
    Javier Duarte$^3$,
    Johannes Haller$^4$,
    Gregor Kasieczka$^4$,
    Artur Lobanov$^4$,
    Vladimir Loncar$^{5,12}$
    Wayne Luk$^2$,
    Jennifer Ngadiuba$^{6},$
    Maurizio Pierini$^{7}$,
    Philipp Rincke$^{8,10}$,
    Arpita Seksaria$^{10}$,
    Sioni Summers$^{7}$,
    Andre Sznajder$^{11}$,
    Alexander Tapper$^{2}$,
    Thea K. \AA rrestad$^1$
}
\address{
$^1$ETH Z\"{u}rich, Z\"{u}rich, Switzerland,\\
$^2$Imperial College London, London, UK,\\
$^3$University of California San Diego, La Jolla, CA, USA,\\
$^4$Universit\"{a}t Hamburg, Hamburg, Germany,\\
$^5$Massachusetts Institute of Technology, Cambridge, MA, USA\\
$^6$Fermi National Accelerator Laboratory, Batavia,  IL, USA,\\
$^7$European Organization for Nuclear Research (CERN), Geneva, Switzerland,\\
$^8$Universit\"{a}t G\"{o}ttingen, G\"ottingen, Germany,\\
$^9$University of Southern California,Los Angeles, CA, USA,\\
$^{10}$Uppsala Universitet, Uppsala, Sweden,\\
$^{11}$Universidade do Estado do Rio de Janeiro (UERJ), Rio de Janeiro, Brazil, \\
$^{12}$Institute of Physics Belgrade, Serbia
}

\ead{podagiu@ethz.ch}

\vspace{10pt}
\begin{indented}
    \item[]\today
\end{indented}

\begin{abstract}
    Three machine learning models are used to perform jet origin classification.
    These models are optimized for deployment on a field-programmable gate array device.
    In this context, we demonstrate how latency and resource consumption scale with the input size and choice of algorithm.
    Moreover, the models proposed here are designed to work on the type of data and under the foreseen conditions at the CERN LHC during its high-luminosity phase.
    Through quantization-aware training and efficient synthetization for a specific field programmable gate array, we show that $\mathcal{O}(100)\units{ns}$ inference of complex architectures such as Deep Sets and Interaction Networks is feasible at a relatively low computational resource cost.
\end{abstract}

\submitto{\MLST}
\maketitle

\section{Introduction}

At the CERN Large Hadron Collider (LHC), proton beams collide every 25\units{ns} in each of the four particle detectors located around the LHC ring.
The collision events generate sprays of outgoing particles that are detected by sensors, which amount to a data rate of tens of terabytes per second.
For the ATLAS~\cite{atlas} and CMS~\cite{CMS} general-purpose experiments, the data throughput is too large to record every single event.
Therefore, a subset of events are selected by a real-time event filtering system, called the \emph{trigger}.

The \textit{current} trigger system consists of two stages.
First, the Level-1 Trigger (L1T) reduces the event rate from $\mathcal{O}(10)$\units{MHz} to $\mathcal{O}(100)$\units{kHz}, rejecting $\sim$99.7\% of all collisions.
The~frequency of collisions and limited buffer size set the maximum L1T latency to~$\mathcal{O}(1)\,\mu$s.
Thus, the L1T is hardware based, with its algorithms running on Field-Programmable Gate Arrays (FPGAs).
The second stage is represented by the High-Level Trigger (HLT).
The HLT consists of software executed on a dedicated CPU farm and further reduces the event rate to 1~kHz.
Only data accepted by the trigger system are saved entirely.
Therefore, a high selection efficiency is of great importance for any LHC measurement and will become even more so after the high-luminosity upgrade.

The LHC will undergo the High-Luminosity (HL-LHC) upgrade between 2026-2028.
The new HL-LHC will provide ten times more data.
This will be achieved by increasing the number of simultaneous interactions per proton collision by a factor of three to~four.
To handle this upcoming increase in data complexity, the particle detectors at the LHC will be upgraded to maintain their detection efficiency for interesting physics processes.
For the CMS experiment, this includes the addition of tracking information to the L1T, which will enable particle-level reconstruction and pileup mitigation as part of the L1T~\cite{phase2}. 
Consequently, Particle-Flow (PF) reconstruction~\cite{PF} will be performed for the first time at the L1T, correlating tracks from the muon and tracking detectors with calorimeter energy clusters to identify each final-state particle in the jet~\cite{phase2}.

Final-state particles originating from the decay and hadronization of initial massive particles, such as top quarks, $W$ bosons, or $Z$ bosons, are clustered into \emph{jets}~\cite{Cacciari:2008gp,Cacciari:2011ma}.
Knowing the particle type from which each jet originated in a collision event could greatly improve the trigger selection algorithms.
This is successfully demonstrated in offline selection algorithms~\cite{jedi,particlenet,whiteson,kasieczka}.
Thus, jet origin identification increases the detector sensitivity for new physics and precision measurements.

Several obstacles must be overcome when designing and deploying such an algorithm.
First, due to the limited amount of resources and time at the L1T, only a small set of particles can be reconstructed and subsequently clustered into jets.
Therefore, there is a limited amount of information available.
Second, particles may arrive \emph{unordered}, since sorting is a resource- and time-intensive operation.
Hence, it would be desirable for a deployed algorithm to be robust against any permutation of the input particles.
Third, individual algorithms must have a maximum latency of $\mathcal{O}(100)\units{ns}$ to be suitable for L1T integration at the HL-LHC.
Furthermore, the system must be able to keep up with the rate of new events, i.e., one every 25\units{ns}, and process up to 10 jets per~event; this~last~constraint is loosened by the use of \emph{Time Multiplexing (TM)}, in which $N_\mathrm{TM}$ processors run identical algorithms on different events~\cite{phase2}.
Fourth, several algorithms run in parallel on each FPGA board, meaning that resources are scarce and individual algorithms should take up significantly less than the total resources available on one~FPGA.
Finally,~the~\mbox{HL-LHC}~L1T selection algorithms must reduce the event rate by a projected six orders of magnitude, compared to the current four, and hence be even more accurate at very low False Positive Rates (FPRs).
To satisfy these challenging requirements, deep neural networks are explored, since this type of algorithms are shown to be relatively fast and accurate in similar classification tasks.

However, conventional machine learning classifiers would not, as they are commonly found in literature, satisfy the latency constraints of the L1T.
Thus, Ref.~\cite{Duarte:2018ite} introduced \hlsfml~\cite{hls4ml}, an open-source Python library for translating machine learning models into FPGA or Application-Specific Integrated Circuit (ASIC) firmware.
Since~there are several L1T algorithms deployed per FPGA, each of them should take only a fraction of the full FPGA resources.
To compress the models, the numerical precision of their the weights and operations are reduced in a process known as quantization~\cite{NIPS2015_5647,han2016deep}.
With~its interface to QKeras~\cite{qkeras}, \hlsfml supports quantization-aware training, making it possible to drastically reduce FPGA resource consumption while preserving accuracy.
Using \hlsfml we can compress neural networks to fit the resources of current FPGAs.

The use of machine learning to classify jets is well-studied for high energy physics and several such algorithms are currently in use in experiments at the LHC.
The most successful such algorithms use the jet constituents as inputs~\cite{jedi,particlenet,efn,parT,garnet,Bogatskiy:2022czk,Bogatskiy:2023nnw,Gong:2022lye}.
Permutation-invariant machine learning algorithms such as Deep Sets (DS)~\cite{ds,efn} and Interaction Network (IN)~\cite{battaglia2,bronstein,graphrev,shlomi}, a type of Graph Neural Network (GNN), are suitable for jet tagging because jet particle data is sparse and has no intrinsic~order.
Additionally, the DS and IN models outperform simpler MLPs when the number of particles is larger than 16.
However, INs are computationally expensive: they apply a Multilayer Perceptron (MLP) to each node and each edge; thus the computational cost scales as $\mathcal{O}(N^2)$, where $N$ is the number of particle constituents of a jet.
In contrast, a DS network applies an MLP to each particle only and thus scales linearly with $N$.

In this work, we implement and compare a variety of exactly permutation-invariant neural networks based on particle-level data, i.e., DS and IN models, as well as MLPs, which are not permutation-invariant.
The MLP, IN, and DS networks we train are designed to have $\mathcal{O}(100)$\units{ns} inference time and synthesized into RTL firmware for an FPGA using \hlsfml and Vivado HLS.
We~show the dependence of these algorithms on $N$ by comparing their classification accuracies, their latencies, throughput, and their resource consumption, as a guide for designing jet origin classifiers for the future trigger systems at the HL-LHC experiments.

The rest of this paper is organized as follows.
Section~\ref{sec:related} discusses related work.
In Section~\ref{sec:data}, we introduce the dataset.
This is followed by a discussion of the model architectures in Section~\ref{sec:models}. 
Further, Section~\ref{sec:quantization} describes how the models are compressed.
We discuss the translation into firmware in Section~\ref{sec:firmware}, before we conclude in Section~\ref{sec:conclusion}.

\subsection{Related Work}
\label{sec:related}
Previous efforts explore tools for translating neural network algorithms into FPGA firmware, as reviewed in Refs.~\cite{nn2fpga,cnn2fpga}.
However, these tools aim at implementations that are not optimized for the L1T systems, and they do not necessarily support the neural network architectures studied here.
Conifer~\cite{conifer} and fwXmachina~\cite{fwXmachina,Carlson:2022dgb,Roche:2023int} feature  custom implementations of boosted decision trees on FPGAs, which achieves the desired L1T constraints, but cannot be extended to neural networks.
LL-GNN~\cite{que2024ll-gnn} proposes a domain-specific low latency hardware architecture for processing GNNs in high energy physics, which involves many manual optimizations.
Our current work leverages some of these manual optimizations and enables an automated design flow with \hlsfml.
\mbox{Nano-PELICAN}~\cite{Bogatskiy:2022czk} is a highly compressed version of PELICAN~\cite{Bogatskiy:2022czk,Bogatskiy:2023nnw}, a permutation- and Lorentz-invariant network.
Moreover, LLPNet~\cite{Bhattacherjee:2023evs} is a lightweight graph autoencoder for tagging long-lived particles in the L1T.
However, FPGA implementations of these models have not yet been studied.
Another long-lived particle trigger is discussed in Ref.~\cite{Coccaro_2023}, featuring latencies compatible with HLT constraints.
Therein, the authors use an approach to model compression that is usually employed in commercial contexts; here~we need manual custom optimizations for our models to satisfy the L1T constraints.

\section{Dataset}
\label{sec:data}
In this work, we analyze the publicly available \hlsfml jet dataset~\cite{dataset150}, consisting of jets stemming from five different origins: light quark ($q$), gluon ($g$), $W$ boson, $Z$ boson, and top quark ($t$), each represented by up to 150 particle constituent four-vectors.
The~constituents are in order of descending transverse momentum, $\ptmomentum$.
The dataset is split into 504,000 (126,000) jets for training (validation) and 240,000 jets for testing, with $k$-folding applied as detailed in Section~\ref{sec:models}.
The initial-state partons and gauge bosons are generated to have $\ptmomentum\approx1\TeV$, while the final-state particle energies and momenta are smeared to achieve CMS-like detector resolutions.
Additional information on the dataset is found in Ref.~\cite{Coleman:2017fiq}.
This dataset does not necessarily reflect the information available in the L1T with utmost accuracy; however, it is adequate for the comparative analysis done in this work.

An average number of 12 constituents is expected for a typical jet in the L1T, while the average number of constituents per jet in the studied data set is 38 due to the high \ptmomentum of the initial-state particles.
Despite this difference, we use the \hlsfml data set as it is a benchmark for this type of application.
The number of constituents per jet is truncated to the first $N$ highest $\ptmomentum$ particles with $\ptmomentum>2\GeV$ and then randomly shuffled; this is done to mimic the HL-LHC L1T scenario where the jet particles would be unordered.
The $2\GeV$ threshold is motivated by the CMS L1T tracking planned for the HL-LHC~\cite{phase2}, which will reconstruct tracks down to 2\GeV.
The $N\in \{8, 16, 32\}$ cases are studied to quantify the effect of $N$ on different model metrics, e.g., accuracy, latency, and resources.
Whenever a jet contains less than $N$ constituents, the data is zero padded up to $N$.

The constituent features we use are the transverse momentum $\ptmomentum$, the pseudorapidity difference relative to the jet axis $\eta_\mathrm{rel}$, and the azimuthal angle relative to the jet axis~$\phi_\mathrm{rel}$.
In contrast with the current L1T, the latter two will be available at the HL-LHC L1T.
As the particle $\ptmomentum$ has significantly higher values than $\eta_\mathrm{rel}$ and $\phi_\mathrm{rel}$, their distributions are normalized with respect to their corresponding [5, 95]\% interquantile range; this method is used instead of the full range of the feature for robustness against data outliers. 
This~process brings $\ptmomentum, \eta_{\mathrm{rel}}, \phi_\mathrm{rel}$ to the same order of magnitude.
Furthermore, this division can be achieved using a bit shift on the FPGA and thus has a negligible impact on the key model metrics.

\section{Model architectures}
\label{sec:models}
The input data consists of $N$ jet constituents, each with the three features $(\ptmomentum, \eta, \phi)$.
For~the IN, each jet is represented as a fully-connected graph, where the graph nodes are the jet constituents defined by the three aforementioned features.
Meanwhile, for the DS, the data is represented as a collection of independent points and the algorithm acts on each point separately.
For the MLP, the constituent dimension of the data is flattened and the network receives a 1D list of values.
The models we use are all 5-class classifiers, implemented using the TensorFlow~\cite{tensorflow} and Keras~\cite{keras} libraries.

\begin{figure}[tpb]
    \centering
    \includegraphics[width=0.9\textwidth]{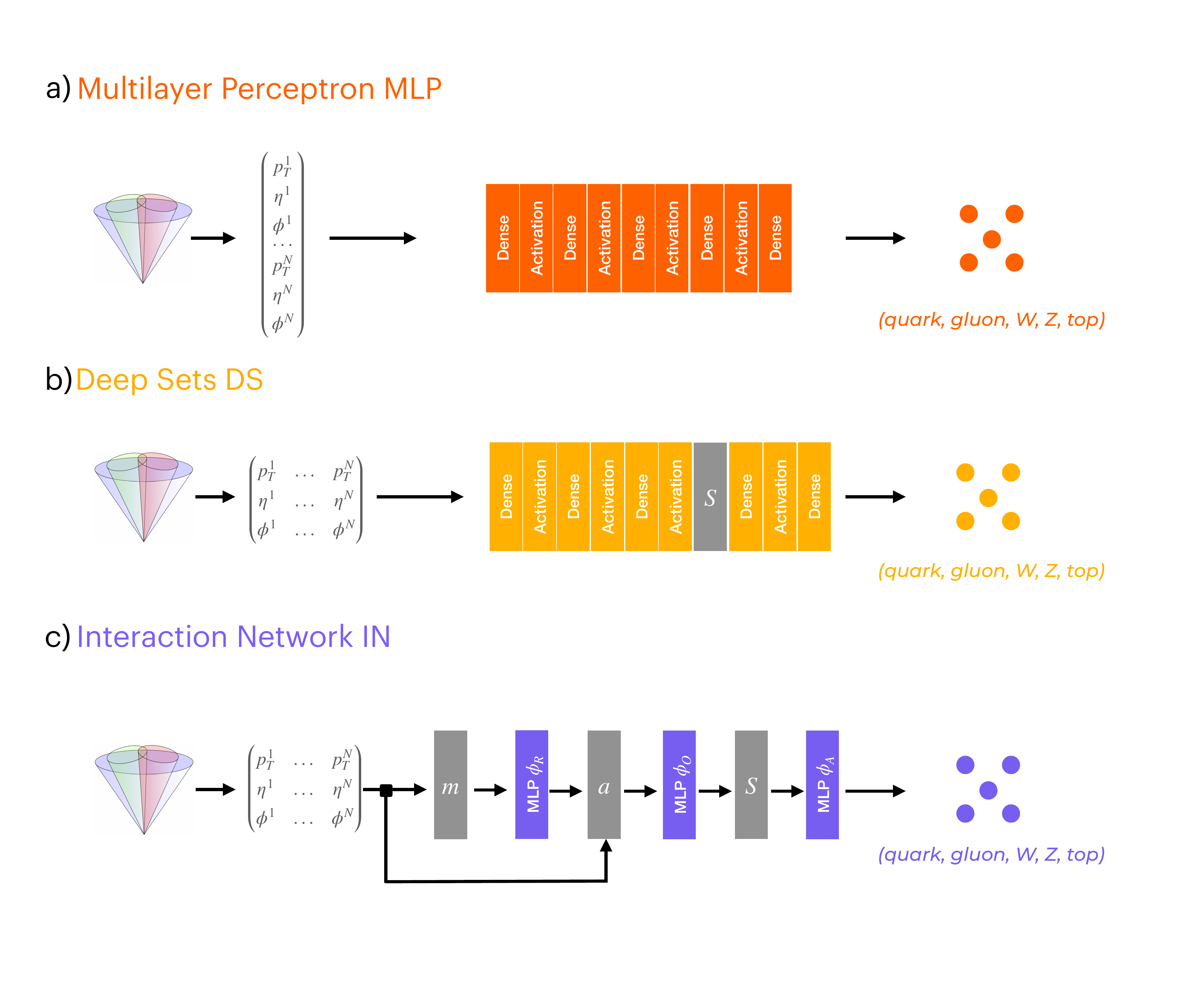}
    \vspace{-0.1cm}
    \caption{Schematic of all the considered models.
        \textbf{(a)} A simple multilayer~perceptron.
        The 2D input is flattened to one dimension before it is passed through the~MLP.
        \textbf{(b)}~A~permutation-invariant deep set network~\cite{ds}.
        The initial MLP acts on the features of each input constituent.
        The aggregation layer $S$ performs a permutation invariant operation at the constituent level and hence brings the input to a 1D vector.
        \textbf{(c)} A permutation-invariant interaction network~\cite{battaglia} as implemented before in Ref.~\cite{jedi}.
        The~input~is transformed by the marshaling function $m$ into a fully-connected graph.
        The exact mechanics of the IN are described in Section~\ref{sec:models}.
        For each network type, all hyperparameters, such as the number of layers and number of nodes per layer, are optimized depending on the number of constituents.
        The hyperparameters of each network are presented in Section~\ref{sec:models} and their respective performance is shown in Table~\ref{table:modelvsNcontituents}.}
    \label{fig:models}
\end{figure}

The output layer of all these models consists of a fully-connected layer with five nodes and a softmax activation function. 
Thus, the model returns the probabilities for a given jet sample to originate from one of the five classes listed at the start of Section~\ref{sec:data}.
Based on the nature of the input data and the strict latency and resource constraints, we explore simple MLP models and permutation-invariant DS and IN models.
The former is considered due to their favorable low latency inference, whereas the latter are expected to have a higher classification accuracy.
We use the following specific architectures:
\begin{table}[tb!]
    \centering
    \caption{
    Floating-point model performance for 8, 16, and 32 jet constituents. 
    The~uncertainties on the AUCs are all $\sim 0.001$ and thus not included for legibility.
    }
    \resizebox{\textwidth}{!}{
        \begin{tabular}{l c c c c c c c c c}
            \multirow{2}{*}{Architecture} & \multirow{2}{*}{Constituents} & \multirow{2}{*}{Parameters} & \multirow{2}{*}{FLOPs} & \multirow{2}{*}{Accuracy} & \multicolumn{5}{c}{AUC}                             \\
                                          &                               &                             &                        &                           & $g$                     & $q$  & $W$  & $Z$  & $t$  \\
            \hline
            MLP                           & \multirow{3}{*}{8}            & 26,826                      & 53,162                 & $64.6 \pm 0.1 \%$         & 0.84                    & 0.88 & 0.90 & 0.88 & 0.92 \\
            DS                            &                               & 3,461                       & 36,805                 & $64.0 \pm 0.3 \%$         & 0.84                    & 0.88 & 0.90 & 0.88 & 0.92 \\
            IN                            &                               & 3,347                       & 37,232                 & $64.9 \pm 0.2 \%$         & 0.84                    & 0.88 & 0.91 & 0.89 & 0.92 \\
            \hline

            MLP                           & \multirow{3}{*}{16}           & 20,245                      & 40,485                 & $68.4 \pm 0.3 \%$         & 0.87                    & 0.89 & 0.91 & 0.90 & 0.94 \\
            DS                            &                               & 3,461                       & 71,109                 & $69.4 \pm 0.2 \%$         & 0.87                    & 0.89 & 0.93 & 0.92 & 0.94 \\
            IN                            &                               & 3,347                       & 140,432                & $70.8 \pm 0.2 \%$         & 0.88                    & 0.90 & 0.94 & 0.92 & 0.94 \\
            \hline

            MLP                           & \multirow{3}*{32}             & 24,101                      & 48,197                 & $66.2  \pm 0.2 \%$        & 0.90                    & 0.89 & 0.89 & 0.88 & 0.94 \\
            DS                            &                               & 3,461                       & 139,717                & $75.9 \pm 0.1 \%$         & 0.91                    & 0.91 & 0.96 & 0.95 & 0.95 \\
            IN                            &                               & 7,400                       & 109,556                & $75.8 \pm 0.3 \%$         & 0.91                    & 0.91 & 0.96 & 0.95 & 0.95 \\
            \hline
        \end{tabular}}
    \label{table:modelvsNcontituents}
\end{table}


\newpage

\begin{enumerate}[label=(\alph*)]
    \item A simple MLP as shown in Figure~\ref{fig:models}(a).
          The number of layers, the number of nodes, and the other hyperparameters of the MLP varies with the number of input jet constituents $N$.
          For 8 constituents, the MLP consists of 8 hidden layers with $\{120, 60, 32, 64, 64, 64, 32, 44\}$ nodes, where we apply L1 regularization throughout with a coefficient of $1.31\times 10^{-5}$; this network is trained with a learning rate of $0.0013$ and batch size of $128$.
          In the $N=16$ case, the MLP has 5 hidden layers with $\{88, 88, 44, 44, 44\}$ nodes and an L1 regularization coefficient of  $2.36\times 10^{-5}$; the 16 constituent MLP is trained with an initial learning rate of $0.0015$ and batch size of $256$.
          Finally, for 32 constituents, the MLP is composed of 7 hidden layers with $\{84, 88, 32, 32, 44, 32, 44\}$ nodes, has an L1  coefficient of $3.14\times 10^{-5}$, and is trained using an initial learning rate of $0.0047$ and a batch size~of~$1024$.
          All these three networks use the ReLU activation function~\cite{relu1,relu2} and the Adam optimizer~\cite{kingma2017adam}, where the learning rate is divided by 10 for every 15 epochs of no accuracy improvement. 
          The training stops when accuracy stagnates for 20 epochs.
    \item A deep set network~\cite{ds} as schematically illustrated in Figure~\ref{fig:models}(b).
          The first MLP $\phi$ of this network acts on the features of each constituent independently, mapping the 3 input features to some output dimension $D$.
          Then, this output $A$ with dimensionality $(N, D)$ is aggregated by $S$ over the constituents $N$, reducing the data from a 2D matrix to a one dimensional vector of length $D$.
          Finally, a second MLP $\rho$ is applied to the aggregation output to produce the jet class predictions.
          For any $N\in \{8, 16, 32\}$, $\phi$ is comprised of 3 hidden layers, each with 32 nodes.
          The~aggregation $S$ is chosen to be an average instead of a maximum, since they give similar results, but computing the number of FLOPs for the average is much more trivial than for the maximum.
          The second MLP $\rho$ uses only one hidden layer with 32 nodes, excluding the output layer.
          ReLU activation is used for all DS~networks.
          The 8 and 16 constituent cases are trained with a batch size of 256 and learning rates of 0.0018  and 0.0029, respectively.
          Meanwhile, the 32 constituent DS is trained with a batch size of 128 and a rate of 0.0032. 
          All~the~DS models use the same optimizer, learning rate decay, and early stopping parameters as the MLP.
    \item An interaction network~\cite{jedi,battaglia} that consists of an edge MLP $\phi_R$, followed by a node MLP $\phi_O$, and a graph  
          classifier MLP $\phi_A$, as shown in Figure~\ref{fig:models}(c).
          The $\phi_R$ network takes input features from a pair of nodes and learns a set of~different~edge~features.
          The~edge features are aggregated at the corresponding receiver nodes, and concatenated with the original node features as input to $\phi_O$.
          The output embeddings are then averaged by $S$ over the $N$ constituents, and given as input to the graph classifier MLP, which consists of a single ReLU activated fully-connected layer, excluding the output layer.
          The $\phi_R$ and $\phi_O$ MLPs are implemented using 1D convolutions of unit kernel size and unit stride, where weights are shared across the edges and~nodes.
          For~8 and 16 constituents, the MLP $\phi_R$ consists of $2$ hidden layers with $\{12, 6\}$~neurons.
          Meanwhile, the $\phi_R$ for 32 constituents has only one hidden layer with three neurons due to the limited hardware resources.
          The~node MLP $\phi_O$ has three hidden layers with $\{36, 18, 6\}$ neurons for all cases.
          The graph classifier MLPs $\phi_A$ have one hidden layer with 170, 170, and 512 neurons for the 8, 16 and 32 constituent models, respectively.
          Both the IN models for 8 and 32 constituents are trained with a batch size of 128, while the batch size is 512 for the 16 constituent model.
          All the IN models use the Adam optimizer with a learning rate of 0.0005 and early stopping after 40 epochs of no accuracy improvement on the validation data.
\end{enumerate}
\newpage
For all the models, Tensorflow~\cite{tensorflow2015-whitepaper} version 2.8 and QKeras~\cite{qkeras} version 0.9 are used.
The~hyperparameter optimization constraints are set such that the model can fit on the Xilinx Virtex UltraScale+ VU13P FPGA.
This specific FPGA is chosen because it is representative of the future HL-LHC L1T hardware platform.
The hyperparameter optimization is performed automatically using Optuna~\cite{akiba2019optuna} for the MLP and DS, while for the IN it is performed using grid search. 
The models presented here are not necessarily the best models that could be achieved with this data, due to hardware~constraints.
However, they are the best possible models that can be \emph{synthesized} on the chosen FPGA device, given the computational limitations of the hyperparameter optimization process and the simple model compression techniques that we consider.
Additionally, pruning is applied to all the 32 constituent MLP, IN, and DS models such that they fit within the resource constraints of the FPGA.
We~prune the 32 constituent models using the TensorFlow Model Optimization Toolkit, with a polynomial decay schedule~\cite{zhu2017prune} and target sparsity of 50\%.
The pruning is done only for the 32 constituent case since the 32 constituent IN is too large given the available resources of the chosen FPGA. 
The~performance of the models is shown in Table~\ref{table:modelvsNcontituents}.
The uncertainty on the AUC and FPR is obtained using $k$-fold cross validation with $k=5$.
The training dataset is split into 5 such that 1/5 is used for validation and the remaining 4/5 is used for training.
The uncertainties on the figures of merit, AUC and FPR, are quantified by the standard deviation across the 5 folds and found to be $\mathcal{O}(0.1)\%$.
The uncertainties due to random initalizations of model parameters are studied as well and found to be negligible.

\begin{figure*}[tb]
    \centering
    \includegraphics[width=0.92\textwidth]{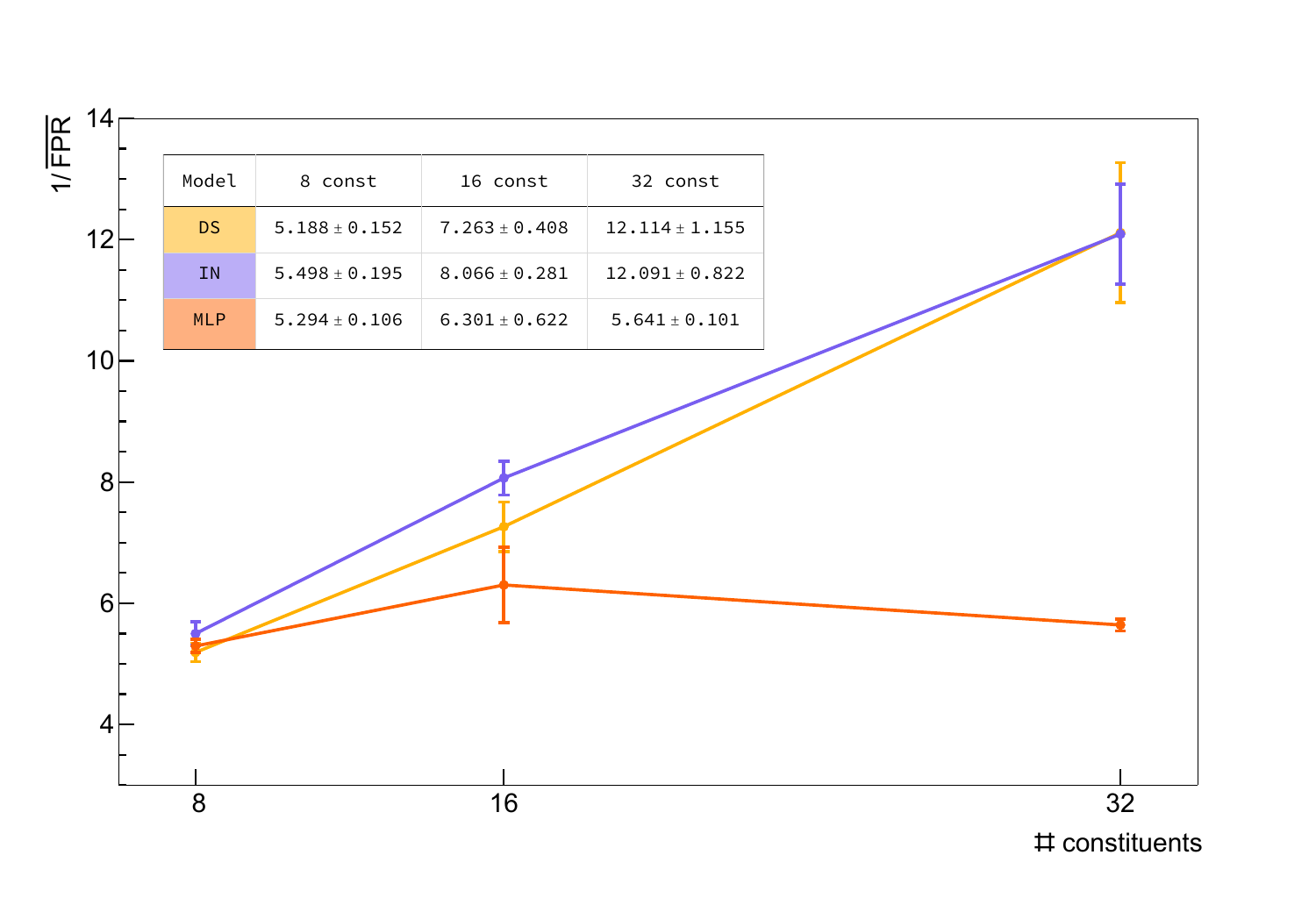}
    \caption{
        The inverse of the Average False Positive Rate ($\overline{\mathrm{FPR}}$) at a fixed true positive rate (TPR) of 80\% over $k=5$ folds of data for $N\in\{8, 16, 32\}$ constituents per~jet.
        This~TPR is chosen since it is a conventional working point in related literature.
        The~size of the MLP is constrained by requiring it to be synthesizable in \hlsfml.
        Therefore, the number of parameters per consecutive layer is limited and the MLP performance decreases from 16 to 32 constituents.
        This is not a factor for the other networks that use a 2D representation of the data.
        The models are quantized to 8 bits.}
    \label{fig:fprvsN}
\end{figure*}

Figure~\ref{fig:fprvsN} shows the inverse of the average FPR across the 5 classes at 80\% TPR, i.e., the inverse average mistagging rate, for each model as a function of input constituents $N$.
The models whose performance is shown in this figure are not the floating point models, but their weights and activations are quantized to 8 bits; moreover, the 32 constituent models are 50\% pruned.
The details are explained in Section~\ref{sec:quantization} and \ref{sec:firmware}.
For now, notice that the models perform similarly if only the highest $\ptmomentum$ 8 constituents are considered.
However, as~the~number of input constituents $N$ increases from 8 to 32, the IN and the DS have a higher $1/\overline{\mathrm{FPR}}$ than the MLP.

In addition, while the mistagging rate decreases significantly for the IN and DS as the number of input constituents increases, the mistagging rate increases for the~MLP.
Increasing the MLP size within the constraints imposed by the High Level Synthesis~(HLS) compiler and the targeted FPGA did not lead to an improvement in the MLP performance.
This effect is most likely due to the lack of ordering in the constituents and to the increase in sparsity with the number of constituents.

This implies that for an L1T system where more than 8 unordered jet constituents are available, using a set or a fully-connected graph representation is beneficial in terms of signal efficiency.
As can be seen from Table~\ref{table:modelvsNcontituents}, this is, however, at the cost of a significantly higher number of floating-point operations necessary for the DS or IN, which implies that the models come at a higher FPGA resource cost.
Ultimately, a trade-off must be made between acceptable signal efficiency and computational resource costs.

\section{Model compression by quantization}
\label{sec:quantization}
We compress the optimized floating-point models by quantization, using QKeras~\cite{Coelho:2020zfu,qkeras}.
The~quantization is performed using the straight-through estimator where layers are quantized during the forward pass, but not for backpropagation.
The models are trained scanning the bit widths from 2 to 16, with the number of integer bits set to~zero.
Furthermore, all parameters are quantized to the same bit width, while the activations are fixed to 8 bits.
The quantized counterpart of each IN architecture is implemented using QKeras supported layers, such as fully-connected and convolutional layers.
Additionally, we developed a custom Keras layer to project the features from the nodes to edges and vice versa through multiplication with the sender or receiver adjacency matrices.

\begin{figure*}[tb]
    \centering
    \includegraphics[width=0.92\textwidth]{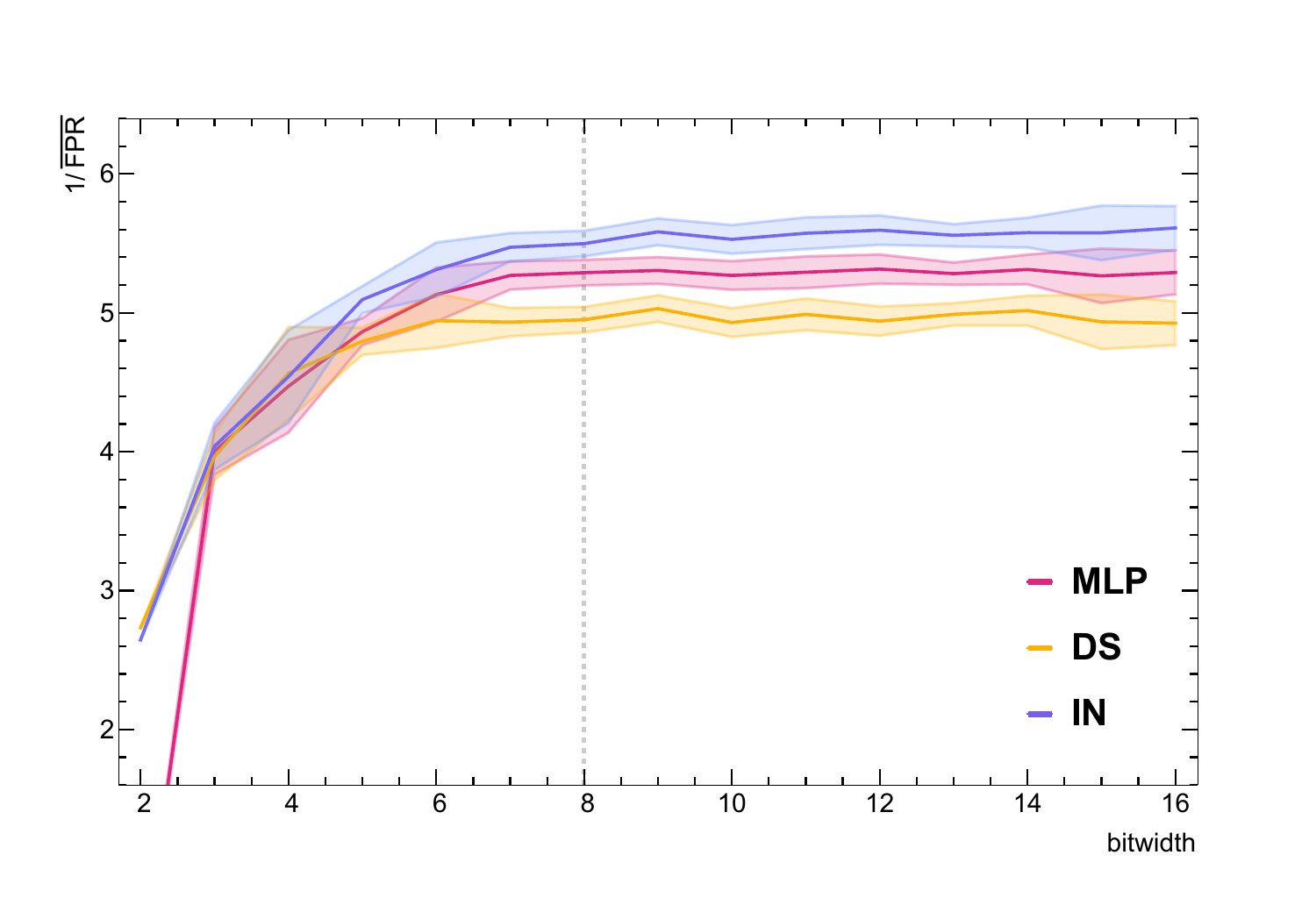}
    \caption{
        The inverted average FPR across the five jet classes, $1/\overline{\mathrm{FPR}}$, as a function of the bit width for the IN, DS, and MLP.
        For each model, the threshold on the classifier score corresponds to a TPR of 80\%.
        The model performance shown here is determined on the $N=8$ data set.
        A bit width of 8 maintains good classification accuracy.
    }
    \label{fig:fpr_bitwidth}
\end{figure*}

The effect of quantization on the inverse false positive rate averaged over classes, 1/$\overline{\mathrm{FPR}}$, at a fixed TPR of 80\% is shown in Figure~\ref{fig:fpr_bitwidth} for the 8 constituent models.
The~uncertainty band is again estimated using $k$-fold cross validation.
The figure shows that 8-bit precision through quantization-aware training is sufficient to compress the models and simultaneously maintain high jet tagging accuracy for our architectures.

\section{Firmware implementation}
\label{sec:firmware}
The quantized models are translated into firmware using \hlsfml, then synthesized with AMD Vivado HLS 2020.1, targeting a Xilinx Virtex UltraScale+ VU13P (\texttt{xcvu13p-flga2577-2-e}) FPGA with a clock frequency of 200\units{MHz}. 
As mentioned in Sec.~\ref{sec:models}, this particular model is chosen since it is representative of the planned hardware for the HL-LHC trigger.
We use a branch of \hlsfml, available at Ref.~\cite{zhiqiang_walkie_que_2024_10553805}.
Except~for the custom IN projection layers, all others are natively supported by~\hlsfml.
For~the projection layer, custom HLS was included using the extension API of \hlsfml.
This custom HLS code is inspired by the optimizations in Ref.~\cite{que2024ll-gnn}.
Since the adjacency matrices are binary and the columns are one-hot encoded the projection calculations are simplified to elementary load and store operations.

We also use a new parallelized implementation of pointwise 1D convolutional~layers.
Each pointwise layer runs an MLP on each jet constituent, requiring a total of $N\times M_\mathrm{in} \times M_\mathrm{out}$ multiplications where $N$ is the number of jet constituents, $M_\mathrm{in}$ is the number of MLP inputs, and $M_\mathrm{out}$ is the number of MLP outputs.
The amount of parallelization is controlled by the Reuse Factor (RF) in \hlsfml, which is used to balance speed with resource consumption.
The RF specifies how many times a multiplier unit is (re)used to compute all the multiplications in a given layer so that only $N\times M_\mathrm{in}\times M_\mathrm{out}/\mathrm{RF}$ total multiplier units are needed.
To avoid a limitation of the HLS compiler on the number of fully unrolled elements within a function call, we split each layer computation into $N/\mathrm{RF}$ separate function calls each using only $M_\mathrm{in} \times M_\mathrm{out}$ multiplier units.

We first evaluate the FPGA latency and resource consumption for the three different architectures at a numerical precision of 4, 6, and 8 bits.
Table~\ref{table:fpgasummary_8const} shows the latency and resource consumption of the quantized models trained on jets with at most 8~constituents.
These results reflect post-logic-synthesis performance by simulating the FPGA on CPU: the models have not been implemented on a physical FPGA.
The estimates also assume minimal I/O overhead, i.e., the data is directly transferred via the bonded I/O pins.
However, in a realistic implementation, an experiment-specific firmware shell would handle the I/O to transfer and process the data from the optical transceivers, thus providing it to the algorithm blocks.
This I/O overhead would be the same for all the algorithms we compare.
The resources on the FPGA are digital signal processors (DSPs), lookup tables (LUTs), block random access memory (BRAM), and also flip-flops (FFs).
The model that is synthesized on the FPGA using \hlsfml achieves 90\% of the accuracy displayed by a model that is compressed in the same ways, but ran directly on CPU.

\begin{table}[tb!]
  \centering
  \caption{
  Average latency, initiation interval (II), and resource consumption for the MLP, DS, and IN models with weights quantized to a bit width of 4, 6, and 8, trained on jet data with a maximum of 8 constituents. 
  The activation functions in these models are quantized to a fixed bit width of 8 to preserve performance.
  The $\mathrm{cc}$ next to the latency and II represents the number of clock cycles on the FPGA. 
  The numbers in parentheses next to the FPGA resource values correspond to the used percentage of the given resource.
  The accuracy ratio between the models presented in this table and the quantized models before FPGA implementation are all above 0.9.
    }
  \resizebox{\textwidth}{!}{
    \begin{tabular}{l c c c c c c c c}
      \multicolumn{8}{l}{FPGA: Xilinx Virtex UltraScale+ VU13P}                                                                                    \\
      Architecture         & Precision & RF & Latency [ns] (cc) & II [ns] (cc) & DSP            & LUT              & FF              & BRAM18      \\
      \hline
      \multirow{3}{*}{MLP} & 4         & 1  & 95 (19)           & 5 (1)        & 101 (0.8\%)    & 235,080 (13.6\%) & 90,150  (2.6\%) & 4 (0.1\%)   \\
                           & 6         & 1  & 95 (19)           & 5 (1)        & 292 (2.4\%)    & 313,371 (18.3\%) & 114,712 (3.3\%) & 4 (0.1\%)   \\
                           & 8         & 1  & 105 (21)          & 5 (1)        & 262 (2.1\%)    & 155,080 (7.6\%)  & 25,714 (0.6\%)  & 4 (0.1\%)   \\
      \hline
      \multirow{3}{*}{DS}  & 4         & 2  & 95 (19)           & 15 (3)       & 101 (0.8\%)    & 235,359 (13.6\%) & 90,190  (2.6\%) & 4  (0.1\%)  \\
                           & 6         & 2  & 95 (19)           & 15 (3)       & 292 (2.4\%)    & 313,230 (18.1\%) & 114,745 (3.3\%) & 4  (0.1\%)  \\
                           & 8         & 2  & 95 (19)           & 15 (3)       & 626 (5.1\%)    & 386,294 (22.3\%) & 121,424 (3.5\%) & 4  (0.1\%)  \\
      \hline
      \multirow{3}{*}{IN}  & 4         & 2  & 150 (30)          & 10 (2)       & 5 (0.0\%)      & 276,720 (16.0\%) & 124,354 (3.6\%) & 12  (0.2\%) \\
                           & 6         & 2  & 155 (31)          & 15 (3)       & 673 (5.5\%)    & 387,625 (22.4\%) & 161,685 (4.7\%) & 12  (0.2\%) \\
                           & 8         & 2  & 160 (32)          & 15 (3)       & 2,191 (17.8\%) & 472,140 (27.3\%) & 191,802 (5.5\%) & 12 (0.2\%)  \\
      \hline
    \end{tabular}}
  \label{table:fpgasummary_8const}
\end{table}


\begin{table}[htb!]
   \centering
   \caption{Number of jet constituents, reuse factor, latency, initialization interval (II) and resource consumption for the models quantized to 8 bits.
      The $\mathrm{cc}$ next to the latency and II represents the number of clock cycles on the FPGA.
      The numbers in parentheses next to the FPGA resource values correspond to the used percentage of the given resource.
      The accuracy ratio between the models presented in this table and the quantized models before FPGA implementation are all above 0.9.
   }

   \resizebox{\textwidth}{!}{
      \begin{threeparttable}
         \begin{tabular}{l c c c c c c c c}
            \multicolumn{8}{l}{FPGA: Xilinx Virtex UltraScale+ VU13P}                                                                                          \\
            Architecture         & Constituents & RF & Latency [ns] (cc) & II [ns] (cc) & DSP            & LUT                & FF               & BRAM18      \\
            \hline

            \multirow{3}{*}{MLP} & 8            & 1  & 105 (21)          & 5 (1)        & 262 (2.1\%)    & 155,080 (9.0\%)    & 25,714 (0.7\%)   & 4 (0.1\%)   \\
                                 & 16           & 1  & 100 (20)          & 5 (1)        & 226 (1.8\%)    & 146,515 (8.5\%)    & 31,426 (0.9\%)   & 4 (0.1\%)   \\
                                 & 32\tnote{a}  & 1  & 105 (21)          & 5 (1)        & 262 (2.1\%)    & 155,080 (7.2\%)    & 25,714 (0.7\%)   & 4 (0.1\%)   \\
            \hline
            %
            %
            \multirow{3}{*}{DS}  & 8            & 2  & 95  (19)          & 15 (3)       & 626 (5.1\%)    & 386,294 (22.3\%)   & 121,424 (3.5\%)  & 4  (0.1\%)  \\
                                 & 16           & 4  & 115 (23)          & 15 (3)       & 555 (4.5\%)    & 747,374 (43.2\%)   & 238,798 (6.9\%)  & 4  (0.1\%)  \\
                                 & 32\tnote{a}  & 8  & 130 (26)          & 10 (2)       & 434 (3.5\%)    & 903,284 (52.3\%)   & 358,754 (10.4\%) & 4  (0.1\%)  \\
            \hline
            \multirow{3}{*}{IN}  & 8            & 2  & 160 (32)          & 15 (3)       & 2,191 (17.8\%) & 472,140 (27.3\%)   & 191,802 (5.5\%)  & 12 (0.2\%)  \\
                                 & 16           & 4  & 180 (36)          & 15 (3)       & 5,362 (43.6\%) & 1,387,923 (80.3\%) & 594,039 (17.2\%) & 52 (1.9\%)  \\
                                 & 32\tnote{a}  & 8  & 205 (41)          & 15 (3)       & 2,120 (17.3\%) & 1,162,104 (67.3\%) & 761,061 (22.0\%) & 132 (2.5\%) \\
            \hline
         \end{tabular}

         \begin{tablenotes} \footnotesize
            \item[a] Pruning to a sparsity of 50\% is applied to the 32-constituent IN model such that it can fit within the resource constraints of the FPGA. For consistency, the same pruning sparsity is applied to the 32-constituent MLP and DS models.
            \normalsize
         \end{tablenotes}

      \end{threeparttable}}
   \label{table:fpgasummary_all}
\end{table}

%


A fully parallel implementation is possible for all MLPs by setting the RF in \hlsfml to 1, such that each network multiplication is distributed across all the resources.
For~the~DS and IN models, the $\mathrm{RF}\in\{2, 4, 8\}$ is set for $N\in\{8, 16, 32\}$ constituents respectively, due to the limited amount of hardware resources.
Increasing the RF reduces the model resource consumption at the cost of increasing its latency and throughput.
Equally important for the throughput is the initiation interval (II), which represents how many clock cycles need to elapse before the network is ready to receive new inputs.
The~II~is higher for the DS and IN models than the MLP, but this can be partially compensated by running several instances of the model in parallel.

\newpage
The way this is accomplished in trigger systems is through time multiplexing, in which $N_\mathrm{TM}$ trigger processor boards run in parallel each processing different events.
For~example, a typical choice is $N_\mathrm{TM}=6$, meaning the II to process an entire event would be $(25\units{ns})(N_\mathrm{TM}) = 150\units{ns}$.
However, each recorded event contains multiple jets.
Assuming that 10 jets are classified sequentially per event, the maximum allowable II per jet would correspond to approximately 15\units{ns}, which is perfectly consistent with Table~\ref{table:fpgasummary_all}.
We note that given the size of the models, this approach may not be feasible.


Table~\ref{table:fpgasummary_all} shows how resource consumption and latency scale as a function of the number of input jet constituents for the three different architectures.
While the latency remains relatively unchanged as the number of constituents increases for the MLP, the latency is proportional to the number of constituents for the DS and IN.
For cases where the number of constituents is large, using a DS or IN architecture is advantageous.
However, from Table~\ref{table:fpgasummary_all}, this incurs additional resources and latency.
One partial solution to this resource problem is to use advanced pruning methods~\cite{optimalbraindamage,han2016deep,zhu2017prune,frankle2019lottery,learningraterewinding,supermask,stateofpruning}, where insignificant weights are removed while the model performance is maintained.
In~this work, we use pruning for the 32 constituent models, although our pruning process is rudimentary and done to fit the IN model into the available resources of the chosen~FPGA.
When the model is synthesized, the pruned weights are set to zero and the corresponding operations are skipped.
Different pruning algorithms~\cite{frankle2019lottery,learningraterewinding,supermask}~might~perform~better.
Alternatively, the reuse factor could be increased to achieve lower resource consumption.
However, this implies higher latencies, which in the L1T context is not worth paying.
Exploration of additional model compression paradigms is left for future work.

\newpage

\section{Conclusion and future work}
\label{sec:conclusion}
Neural network based jet classification algorithms are synthesized on FPGA devices that mimic the environment within the hardware layers of the real-time data processing systems for a typical LHC experiment after the high-luminosity upgrade.
Using jet data with constituent level information, we show how one could synthesize machine learning algorithms pertaining to three different data representations on an FPGA by using the \hlsfml library.
We also demonstrate how metrics like accuracy, latency, and resource, utilization scale as a function of the number of input jet constituents: an improvement in accuracy is gained by using a set or fully-connected graph representation when the number of jet constituents is larger than 8. Meanwhile, the Deep Sets network strikes a good balance between accuracy, latency, and resource consumption compared with the deployed and tested MLP and IN models.
Employing quantization-aware training and, for the 32 constituent case, pruning, we show how to efficiently limit resource utilization of these models while retaining accuracy.

In conclusion, we have identified and shown the necessary ingredients to deploy a jet classifier in the level-1 trigger of the high-luminosity LHC experiments, when high-granularity particle information and particle-flow reconstruction would be accessible.
An algorithm of this kind could significantly improve the quality of the trigger decision and improve signal acceptance, increasing the scientific reach of the experiments.
Additionally, the results shown in this work could be improved upon by employing advanced model compression techniques, more thorough hyperparameter optimization, and better synthesis fine-tuning.
Moreover, the presented results, although representative, are from post-synthesis but pre-implementation algorithms.
A full FPGA implementation is also left for future work.

\section{Data availability}
The data used in this study are openly available at Zenodo at Ref.~\cite{dataset150} under DOI 10.5281/zenodo.3602260.
The software used in this study is also available at Zenodo at Ref.~\cite{zhiqiang_walkie_que_2024_10553805} under DOI 10.5281/zenodo.10553804.

\section{Author information}

\subsection{Corresponding author}
Correspondence and material requests can be emailed to P. Odagiu (\href{mailto:podagiu@ethz.ch}{podagiu@ethz.ch}).

\newpage
\ack
P.O. and T.\AA. are supported by the Swiss National Science Foundation Grant No.~PZ00P2\_201594.
M.~P. and V.~L. are supported by the European Research Council (ERC) under the European Union's Horizon 2020 research and innovation program (grant agreement n$^o$ 772369).
M.~P., V.~L., and S.~S. are partially supported by the European Research Council (ERC) under the European Union's Horizon 2020 research and innovation program (grant agreement n$^o$ 966696).
J.~D. is supported by the U.S. Department of Energy (DOE), Office of Science, Office of High Energy Physics Early Career Research program under Award No. DE-SC0021187 and the NSF under Cooperative Agreement OAC-2117997 (A3D3 Institute).
A.~S. is supported by the following Brazilian research agencies: CAPES, CNPq, and FAPERJ. The members of the Hamburg University group acknowledge the support of the Deutsche Forschungsgemeinschaft (DFG, German Research Foundation) under Germany’s Excellence Strategy---EXC 2121 ``Quantum Universe''---390833306. J.~N. is supported by Fermi Research Alliance, LLC under Contract No. DE-AC02-07CH11359 with the Department of Energy (DOE), Office of Science, Office of High Energy Physics.
J.~N. is also supported by the U.S. Department of Energy (DOE), Office of Science, Office of High Energy Physics ``Designing efficient edge AI with physics phenomena'' Project (DE-FOA-0002705).
Z.Q. and W.L. are supported by the United Kingdom EPSRC (grant numbers EP/V028251/1, EP/L016796/1, EP/N031768/1, EP/P010040/1, and EP/S030069/1).

\newpage


\bibliographystyle{cms_unsrt}
\bibliography{main}

\end{document}